\documentclass[final,5p,times,twocolumn]{elsarticle}

\usepackage{amssymb}
\usepackage{amsmath}
\usepackage{amsfonts}
\usepackage{stackengine}
\usepackage{comment}
\usepackage{graphicx}
\usepackage{braket}
\usepackage{calligra}
\usepackage{xcolor}
\DeclareMathAlphabet{\mathcalligra}{T1}{calligra}{m}{n}
\newcommand{\ag}{\textcolor{black}}
\DeclareMathOperator{\Tr}{Tr}

\begin{document}

\begin{frontmatter}



\title{Conditional mutual information: A generalization of causal inference in quantum systems} 
\author[label1]{Anupam Ghosh}
\ead{ghosh@cs.cas.cz}
\affiliation[label1]{organization={Department of Complex Systems},
            addressline={Institute of Computer Science of the Czech Academy of Sciences}, 
            city={Prague},
            postcode={18200}, 
            country={Czech Republic}}

\begin{abstract}
The concept of causality is fundamental to numerous scientific explanations; however, its extension to the quantum regime has yet to be rigorously explored. This letter introduces the development of a quantum causal index, a novel extension of the classical causal inference framework, tailored to learn the causal relationships inherent in quantum systems. Our study focuses on the asymmetric quantum conditional mutual information (QCMI), incorporating the von Neumann entropy, as a directional metric of causal influence in quantum many-body systems.  We analyze spin chains using the QCMI, implementing a projective measurement on one site as the intervention and monitoring its effect on a distant site conditioned on intermediate spins. Additionally, we study the effective causal propagation velocity, which is the speed at which QCMI becomes significant at distant sites. These findings indicate the presence of finite-speed propagation of causal influence, along with the emergence of coherent oscillations.
\end{abstract}




\end{frontmatter}

\section{Introduction}
\label{intro}
The `cause-effect’ relationship between the variables of dynamical systems remains one of the fundamental questions in the literature for a long time~\cite{pearl09,spirtes11,hunter13,kutach13}. Establishing a causal inference, which distinguishes correlation from actual influence, is essential for developing more accurate models and effective interventions. Therefore, understanding causal relationships is crucial for studying dynamical systems, as it clarifies how dynamic states evolve and how various components of a system interact. A fundamental assumption in causal reasoning is based on the `Reichenbach Principle of Common Cause', which philosopher Hans Reichenbach originally articulated in his seminal work, \emph{The Direction of Time}, published in $1956$~\cite{reichenbach56}. Following this principle, if $X$ and $Y$ are two statistically correlated variables and neither $X$ causes $Y$ nor $Y$ causes $X$, then there exists a third variable $Z$, referred to as the \emph{common cause}, that explains this correlation. 
In classical systems, causal relationships are typically formalized using tools such as directed acyclic graphs and structural equation models~\cite{pearl09,spirtes11,hunter13,kutach13}. Most importantly, these classical causal models assume an underlying temporal and spatial ordering of events, consistent with classical intuitions about locality and determinism. However, the transition to quantum mechanics introduces fundamental conceptual and operational deviations from this picture~\cite{riggs09}. Quantum phenomena such as superposition, entanglement, and the no-signaling constraint undermine classical assumptions about causal separability~\cite{jaeger13}. More strikingly, quantum theory admits scenarios where the causal order between events is not well-defined, as exemplified by process matrices exhibiting \emph{indefinite causal structure} \cite{oreshkov12}. These developments suggest that classical causal notions are inadequate for fully capturing the structure of quantum correlations.
The research conducted by Tucci~\cite{tucci95} signifies a pioneering effort to provide a quantum generalization of causal inference. This work~\cite{tucci95} presents a novel perspective of causal dependencies using complex transition amplitudes rather than traditional quantum channels. Further advancements in the formulation of process matrices~\cite{oreshkov12}, quantum Bayesian networks~\cite{leifer2013towards}, and causal modeling in quantum theory~\cite{allen2017quantum} highlight a growing interest and urgency in establishing a rigorous and operational framework for quantum causality. The indices associated with quantum causality are essential not only for enriching our fundamental understanding of quantum mechanics but also for enabling practical applications in quantum computing and quantum communication. For example, quantum causal analyses present an opportunity to design advanced methodologies for simulating many-body systems in condensed matter physics~\cite{leifer08}. Additionally, these analyses may facilitate innovative approaches for discerning the underlying causal structures derived from quantum correlations~\cite{ried15}. Consequently, developing and refining these causal indices is critical for ensuring the conceptual coherence of quantum theory and promoting the advancement of quantum technologies.
In $2001$, Palu\v{s} et al.~\cite{palus01} proposed a specific form of conditional mutual information (CMI), $I_c(A; B|C)$, which is asymmetric in variables $A$ and $B$ and can be used to measure the direction of information flow from $A$ to $B$ when the third variable $C$ is given. The subscript `$c$' in $I_c(A; B|C)$ implies that the discussion associated with  $I_c(A; B|C)$ is restricted to the classical domain. Generally, in information theory~\cite{cover06}, the standard form of CMI $I_c(A:B|C)$ is used to measure the dependency between two variables $A$ and $B$ given the knowledge of the other variable $C$, and $I_c(A:B|C)$ is symmetric in $A$ and $B$. It is essential to clarify that if we intend to utilize CMI as a causal index, it must exhibit asymmetry in its relationship with variables $A$ and $B$. This requirement arises from the fundamental nature of causal indices, which are inherently asymmetric, as they seek to identify and illustrate directional relationships between variables~\cite{pearl09}. In other words, a causal index is designed to uncover directed dependencies between two variables, often by conditioning on third variables or leveraging temporal information. The  causal index $I_c(A; B|C)$, proposed by Palu\v{s} et al.~\cite{palus01}, maintains this asymmetry in $A$ and $B$. Later, Palu\v{s} and Vejmelka~\cite{palus07} reported an updated and higher-dimensional version of CMI, which is a frequently used form of CMI in dynamics and causality studies. A brief outline of this particular form of CMI is discussed in \ref{sec:ccmi}. Existing literature~\cite{palus14,palus24,ghosh25} supports that CMI $I_c(A; B|C)$ has been widely explored in the study of causal structure. Motivated by this asymmetric form of CMI, in this letter, we intend to extend it to study causal relations in quantum systems. More explicitly, \emph{we are interested in reporting a quantum version of CMI, $I(A; B|C)$, which exhibits asymmetry in variables $A$ and $B$ and is suitable for studying causal structure in quantum systems}.
The letter is organized as follows: The detailed formalism of the quantum conditional mutual information (QCMI) is discussed in Sec.~\ref{sec:qcmi}. Subsequently, we discuss the advantage of using asymmetric QCMI over the symmetric QCMI using a GHZ-like state in Sec.~\ref{sec:advantage}. After that, we use QCMI as a causal index to study the spin chains and elaborately discuss it in Sec.~\ref{sec:spin}. Finally, we conclude the main findings in Sec.~\ref{sec:con}.

\section{Quantum conditional mutual information (QCMI)}
\label{sec:qcmi}
Let us consider a tripartite quantum state with the density matrix $\rho_{ABC}$. The general form of QCMI is defined as~\cite{nielsen10}:
\begin{equation}
	\label{eq:qcmi}	
	I(A:B|C) = S(AC) + S(BC) - S(C) - S(ABC),
\end{equation}
where $S(\cdot)$ denotes the von Neumann entropy~\cite{nielsen10}, and it is defined as follows:
\begin{equation}
	S(\rho_{ABC}) = -  \Tr \left[ \rho_{ABC} \log_2 \left( \rho_{ABC}\right) \right].
\end{equation}
The QCMI $I(A:B|C)$ measures how much information subsystems $A$ and $B$ share, {given} access to $C$. In other words, it quantifies the residual correlation between $A$ and $B$ that is {not} explained by $C$. Note that this standard QCMI is symmetric in $A$ and $B$, i.e., 
\begin{equation}
	\label{eq:qcmi_sym}
	I(A:B|C) = I(B:A|C).
\end{equation}
In order to incorporate a \emph{causal direction} --- for example, $A \to B$ --- consider applying a quantum instrument (generalized measurement) on subsystem $A$ before evaluating correlations with $B$. A quantum instrument $\mathcal{M}_A$ is described by a set of measurement operators $\{ M_x \}$ acting on $A$ such that
\begin{equation}
	\sum_x M_x^\dagger M_x = I_A,
\end{equation}
where $I_A$ is the identity operator acting on the Hilbert space of subsystem $A$, and the post-measurement state (on $BC$) given outcome $x$ is
\begin{equation}
	\rho_{BC}^x = \frac{1}{p_x} \mathrm{Tr}_A \left[ (M_x \otimes I_{BC}) \rho_{ABC} (M_x^\dagger \otimes I_{BC}) \right],
\end{equation}
with
\begin{equation}
	p_x = \mathrm{Tr}\left[(M_x \otimes I_{BC}) \rho_{ABC} (M_x^\dagger \otimes I_{BC}) \right].
\end{equation}
Finally, the \emph{asymmetric} QCMI can then be defined as:
\begin{eqnarray}
	\label{eq:qcmi_final}
	I(A;B|C) := && I(A;B|C)_{\rho, \mathcal{M}} \nonumber \\
	= && \sum_x p_x \, I(B : C)_{\rho^x} \nonumber \\
	= && \sum_x p_x \big[ S(\rho_B^x) + S(\rho_C^x) - S(\rho_{BC}^x) \big],
\end{eqnarray}
where $\rho_B^x = \mathrm{Tr}_C(\rho_{BC}^x)$, and it is called the partial trace over the subsystem $C$. Likewise, $\rho_C^x = \mathrm{Tr}_B(\rho_{BC}^x)$. This quantity $I(A;B|C)$ captures how much information about $B$ and $C$ is retained {after a measurement on $A$}, thus encoding a directionality from $A$ to $B$ mediated through $C$. \ag{Although the final expression for QCMI presented in Eq.~\ref{eq:qcmi_final} appears formally symmetric in variables $B$ and $C$, a distinct operational difference, however, arises from the specific causal question being addressed. When evaluating whether interventions on $A$ have an influence on $B$, $B$ is designated as the target variable, while $C$ serves as the conditioning subsystem. Conversely, to investigate the influence from $A$ to $C$, one would compute $I(A;C|B)$.} In contrast to Eq.~\ref{eq:qcmi}, Eq.~\ref{eq:qcmi_final} demonstrates asymmetry between the variables $A$ and $B$. As a result, we have adopted a different notation, $I(A;B|C)$, in Eq.~\ref{eq:qcmi_final} --- rather than using $I(A:B|C)$. 
\ag{In determining the relationship between QCMI and CCMI (An overview of the CCMI can be found in \ref{sec:ccmi}), we recall the following explicit form of CCMI used in classical causal inference:
\begin{equation}
	I_c(A;B|C)
	= I_c(x_j; x'_{j+\tau} \mid \{x'_j, x'_{j-\eta_0}, x'_{j-2\eta_0}\}),
\end{equation}
which quantifies the predictive information flow from a variable $x_j$ to a future variable $x'_{j+\tau}$ beyond what is contained in the past history of $B$. A positive value of $I_c(A;B|C)$ indicates a directed causal influence from $A$ to $B$ conditioned on $C$.}
\ag{Intuitively, to get the analogy, both quantities $I_c(A;B|C)$ and $I(A;B|C)$ measure directed dependence from $A$ to $B$ with conditioning on past variables $\{x'_j, x'_{j-\eta_0}, x'_{j-2\eta_0}\}$ corresponding to conditioning on subsystem $C$. Temporal asymmetry in the classical case is replaced by measurement back-action in the quantum case. If $\rho_{ABC}$ is diagonal in a classical basis and $\mathcal{M}_A$ measures in that basis,
\begin{equation}
	I(A;B|C)
	\longrightarrow
	I_c(x_j; x'_{j+\tau} \mid \{x'_j, x'_{j-\eta_0}, x'_{j-2\eta_0}\}),
\end{equation}
so the QCMI reduces exactly to the CCMI. Detailed steps to reach CCMI from QCMI  have been incorporated in \ref{sec:reduction}. Hence, the proposed QCMI provides a natural quantum generalization of classical information flow measures used in causal inference for dynamical systems.}
\section{Advantage of using asymmetric QCMI for Causal Analysis}
\label{sec:advantage}
A GHZ-like state is an example of maximally entangled quantum state involving three or more subsystems. Consider the three-qubit GHZ-like state:
\begin{equation}
	\ket{\Psi} = \frac{1}{\sqrt{2}}\big( \ket{000} + \ket{111} \big),
\end{equation}
where $\ket{abc} = \ket{a}_A \otimes \ket{b}_B \otimes \ket{c}_C$. The corresponding density matrix is
\begin{equation}
	\rho_{ABC} = \ket{\Psi}\bra{\Psi}.
\end{equation}
Let $\mathcal{M}_A$ be the projective measurement in the computational basis on $A$, with projectors:
\begin{equation}
	M_0 = \ket{0}\bra{0} \quad \text{and} \quad M_1 = \ket{1}\bra{1}.
\end{equation}
The post-measurement probabilities $p_x$ are given by:
\begin{subequations}
	\begin{eqnarray}
		p_0 = \mathrm{Tr}[(M_0 \otimes I_{BC}) \rho_{ABC}] = \frac{1}{2},\\
		p_1 = \mathrm{Tr}[(M_1 \otimes I_{BC}) \rho_{ABC}] = \frac{1}{2},
	\end{eqnarray}
\end{subequations}
and the post-measurement states on $BC$ are
\begin{subequations}
	\begin{eqnarray}
		\rho_{BC}^0 = \frac{1}{p_0} \mathrm{Tr}_A[(M_0 \otimes I_{BC}) \rho_{ABC}] = \ket{00}\bra{00},\\
		\rho_{BC}^1 = \frac{1}{p_1} \mathrm{Tr}_A[(M_1 \otimes I_{BC}) \rho_{ABC}] = \ket{11}\bra{11}.
	\end{eqnarray}
\end{subequations}

All pure states
\begin{equation}
	\rho_{BC}^0 = \ket{00}\bra{00}, \quad \rho_B^0 = \ket{0}\bra{0}, \quad \text{and} \quad \rho_C^0 = \ket{0}\bra{0},
\end{equation}
have zero entropy, i.e.,
\begin{equation}
	S(\rho_{BC}^0) = S(\rho_B^0) = S(\rho_C^0) = 0,
\end{equation}
which further yields
\begin{equation}
	I(B:C)_{\rho^0} = 0.
\end{equation}
Similarly, for $x=1$, we can write
\begin{equation}
	I(B:C)_{\rho^1} = 0.
\end{equation}
Finally, the asymmetric QCMI for this three-qubit GHZ-like state is given by
\begin{equation}
	I(A ; B | C) = \sum_x p_x I(B:C)_{\rho^x} = 0.
\end{equation}
Now, we are interested in comparing $I(A ; B | C)$ with the symmetric QCMI. Before the measurement, we have
\begin{equation}
	I(A:B|C) = S(\rho_{AC}) + S(\rho_{BC}) - S(\rho_C) - S(\rho_{ABC}).
\end{equation}

The GHZ state has:
\begin{subequations}
	\begin{eqnarray}
		&&S(\rho_{ABC})=0,\\
		&&S(\rho_C)=1,\\
		&&S(\rho_{AC})=1,\\
		&&S(\rho_{BC})=1,
	\end{eqnarray}
\end{subequations}
thus
\begin{equation}
	I(A:B|C) = 1.
\end{equation}
This shows the symmetric QCMI $I(A:B|C)$ is {nonzero}, while the asymmetric one $I(A;B|C)$ {vanishes} after measurement, highlighting the asymmetry and causal interpretation. Applying a quantum measurement on $A$ breaks this symmetry, and the \emph{asymmetric} QCMI $I(A;B|C)$ quantifies information flow from $A$ to $B$ conditioned on $C$. In the GHZ example, measuring $A$ collapses correlations, yielding zero asymmetric CMI, contrasting with the non-zero symmetric QCMI. The usual QCMI is symmetric and does not capture causal direction. 
\section{Application of Asymmetric QCMI for Causal Analysis in Spin Chains}
\label{sec:spin}
\begin{figure}[h]
	\centering
	\includegraphics[width=50cm,height=4.5cm, keepaspectratio]{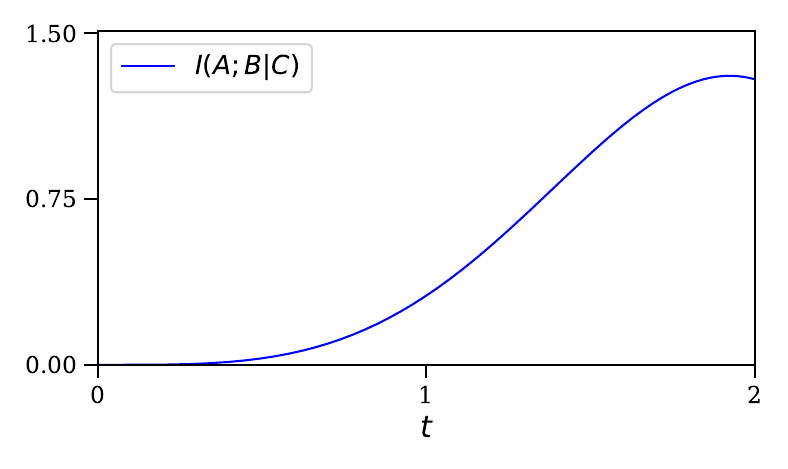}
	\caption{The asymmetric QCMI $I(A;B|C)$ has been plotted as a function of time ($t$) for the four-site XX spin chain. The corresponding parameters are: $N=4$, $J = 1.0$, and $h = 0.5$ in Eq.~\ref{eq:ham_xx}.} 
	\label{fig:four_spins}
\end{figure}

\begin{figure}[h]
	\centering
	\includegraphics[width=50cm,height=4.5cm, keepaspectratio]{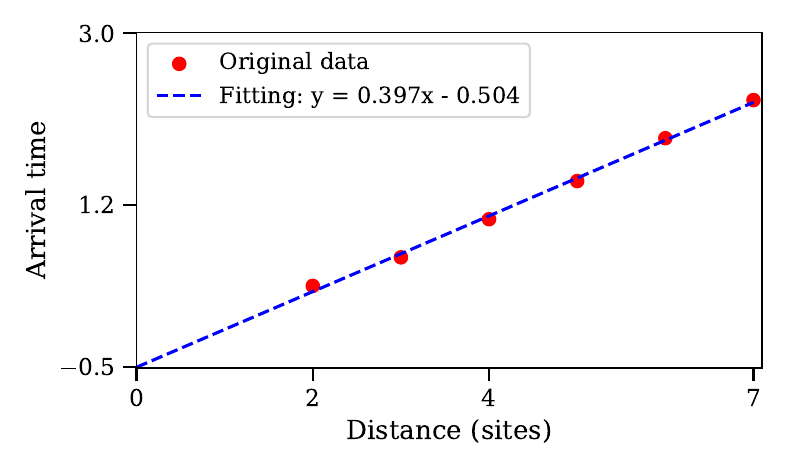}
	\caption{The arrival time $t_\mathrm{arr}(d)$ has been plotted as a function of distance $d$. The effective causal propagation velocity $v_{\mathrm{eff}}$ can be calculated from the slope of this plot using Eq.~\ref{eq:slope}. The associated parameters in Eq.~\ref{eq:ham_xx} are: $N=8$, $J = 1.0$, and $h = 0.3$.} 
	\label{fig:effect}
\end{figure}

Quantum spin chains consist of a sequence of spins (qubits or higher-dimensional spins) arranged linearly, governed by some Hamiltonian. Such systems are key models for exploring many-body quantum physics, entanglement dynamics, and information propagation~\cite{lami25}. Consider a spin chain with sites labeled $1, 2, \ldots, N$ and each site is described by Hilbert space $\mathcal{H}_i$. We further define three subsystems: $A$ represents spin at site $i$, $B$ represents spin at site $j$, and $C$ is some intermediate spins. Intuitively, by applying local operations on one site (site $i$) and measuring the resulting information shared with the other site (site $j$) conditioned on intermediate subsystems (sites between $i$ and $j$), we can obtain a quantitative, directional measure of quantum causal influence that respects the dynamical constraints of many-body quantum systems.
Similar to the previous example, we introduce a quantum instrument (measurement or quantum channel) $\mathcal{M}_A$ on site $i$ and define the asymmetric QCMI:
\begin{eqnarray}
	I(A; B|C) &=& I(A; B|C)_{\rho, \mathcal{M}} \nonumber\\ &=& \sum_x p_x \, I(B : C)_{\rho^x},
\end{eqnarray}
where
\begin{subequations}
	\begin{eqnarray}
		&& \rho^x_{BC} = \frac{1}{p_x} \mathrm{Tr}_A \big[ (M_x \otimes I_{BC}) \rho_{ABC} (M_x^\dagger \otimes I_{BC}) \big], \\ && p_x = \mathrm{Tr}[(M_x \otimes I_{BC}) \rho_{ABC}].
	\end{eqnarray}
\end{subequations}
The joint density matrix $\rho_{BC}$ is obtained from the full chain state $\rho_{ABC}$ by partial tracing. This quantity $I(A; B|C)$ measures how much information about $B$ and $C$ remains after performing $\mathcal{M}_A$ on $A$, revealing directional information flow from $A$ to $B$ mediated by $C$. A positive value of $I(A;B|C)$ suggests that \emph{information injected or measured at site $i$ significantly influences site $j$}, reflecting a causal effect or influence pathway. One can map out the causal network by scanning pairs $(i,j)$ along the chain and varying $C$ (e.g., the spins between $i$ and $j$).
Consider a 1D chain of $N$ sites and each site holds a spin-$\frac{1}{2}$ particle. This spin-$\frac{1}{2}$ chain initialized in a ground state $\rho$ of a nearest-neighbor Hamiltonian, e.g., the transverse-field Ising model:
\begin{equation}
	\label{eq:ham_three}
	H = -J \sum_{k=1}^{N-1} \sigma_z^k \sigma^{k+1}_z - h \sum_{k=1}^N \sigma_x^k,
\end{equation}
with Pauli matrices $\sigma_{x,y,z}^k$ at site $k$. The first term of Eq.~\ref{eq:ham_three} represents the nearest-neighbor Ising interaction with coupling strength $J$, and the second term is associated with the transverse magnetic field with the field strength $h$. Afterwards, we perform a local quench or measurement on site $i$ at time $t=0$, represented by a projective measurement $\mathcal{M}_A$ on spin $i$. Let the system evolve unitarily under $U(t) = e^{-iHt}$ to time $t$. Furthermore, we construct the state $\rho_{ABC}(t)$ for subsystems $A$, $B$, and $C$. We recall that $A$ represents the site $i$, $B$ represents the site $j$, and $C$ represents the intervening spins $\{i+1, \ldots, j-1\}$. Finally, we can compute the asymmetric QCMI $I(A;B|C)$. 
\begin{figure}[h]
	\centering
	\includegraphics[width=50cm,height=4.5cm, keepaspectratio]{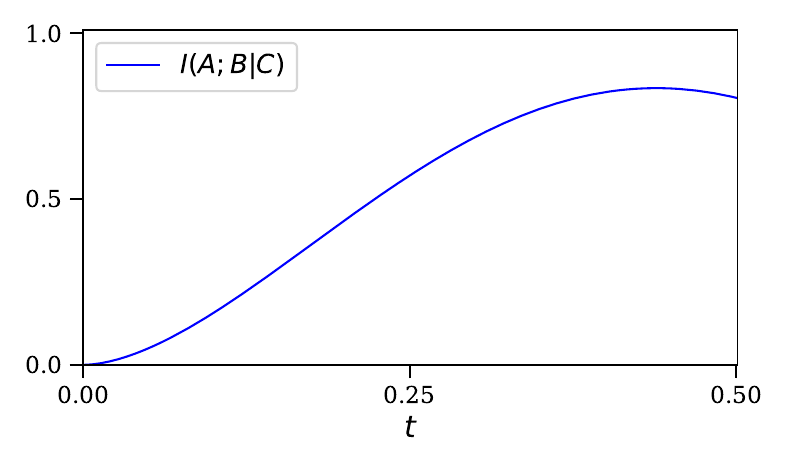}
	\caption{The asymmetric QCMI $I(A;B|C)$ has been calculated for the $3$-qubit spin chain, and it is plotted as a function of time ($t$). The corresponding parameters are: $N=3$ and $J = h = 1.0$ in Eq.~\ref{eq:ham_three}.} 
	\label{fig:three_spins}
\end{figure}
Figure~\ref{fig:three_spins} depicts the variation of $I(A;B|C)$ as a function of time for the $3$-qubit spin chain (i.e., $N=3$ in Eq.~\ref{eq:ham_three}). Here, $i = 0$ and $j=2$. When $t=0$, causal influence beyond nearest neighbors is limited, so $I(A;B|C)$ is negligible for the large value of $|j - i| = 2$. As time progresses, $I(A;B|C)$ increases for spins further away, revealing the spread of causal influence. Thus, the discussed asymmetric QCMI provides a powerful tool for probing causal relations in quantum spin chains. 
Subsequently, we consider a XX spin chain with the following Hamiltonian:
\begin{equation}
	\label{eq:ham_xx}
	H = \frac{J}{2} \sum_{i=0}^{N-2} \left( \sigma_x^{(i)} \sigma_x^{(i+1)} + \sigma_y^{(i)} \sigma_y^{(i+1)} \right)
	+ h \sum_{i=0}^{N-1} \sigma_z^{(i)},
\end{equation}
where $\sigma_\alpha^{(i)}$ denotes the Pauli operator $\sigma_\alpha$ acting on site $i$, $J$ is the nearest--neighbor coupling, and $h$ the transverse field. We initially consider $N=4$, $J=1.0$, and $h=0.5$. Additionally, we adopt the initial state $\ket{1 0 0 0}$ in computational basis. For this example, the intervention is on site $0$, which represents the subsystem $A$. Site $3$ represents the target subsystem $B$, with the conditioning subsystem $C$ as sites $1$ and $2$. Similar to the previous example, Fig.~\ref{fig:four_spins} depicts that $I(A;B|C)$ initially has a value near zero, which further implies that no immediate influence at the distant site $B$. Growth in $I(A;B|C)$ value is visible after a finite delay, which is consistent with the finite-speed propagation of
quantum correlations.
In addition, we are interested in the speed at which QCMI becomes significant at distant sites, which provides an effective causal propagation velocity, and is related to the Lieb--Robinson (LR) bound~\cite{lieb72}. We are interested in asking when a specific intervention will start to matter at a distant site $B$, once an intervention occurred at site $A$ at time $t=0$, and we condition it on intermediate sites $C$. The asymmetric QCMI measures how much the intervention on $A$ changes the correlations between $B$ and $C$ at time $t$. If the intervention has no influence yet on the reduced state of $BC$, then the post-measurement distribution over $BC$ is (almost) unchanged and $I(A;B|C)$ stays near zero. When the effect of the intervention reaches $B$ (through interactions that propagate outwards), the reduced state on $BC$ changes appreciably and $I(A;B|C)$ rises. The time $t_{\rm arr} (d)$ when $I(A;B|C)$ first crosses some small threshold for distance $d=\text{dist}(A,B)$ is therefore an operational `arrival time' for causal influence. The effective velocity is then estimated as
\begin{equation}
	v_{\rm eff} \approx \frac{d}{t_{\rm arr} (d)}. 
\end{equation}
The LR bound makes the above intuition robust. It guarantees no signal can propagate faster than that exponential light-cone. The LR bound gives an exponential suppression of the effect of a local perturbation outside a causal cone of slope $v_{\mathrm{LR}}$, where $v_{\mathrm{LR}}$ is the LR velocity. Therefore, QCMI remains exponentially small outside that cone and only becomes significant when $t$ is large enough that the exponential factor ceases to be tiny. Hence, $v_{\mathrm{LR}}$ is a theoretical upper bound on propagation speed. It depends only on the Hamiltonian (coupling strengths and geometric locality) and is state-independent. \ref{sec:lr} describes analytically and concisely how to calculate the LR velocity for the XX chain (Eq.~\ref{eq:ham_xx}). For the parameter choice used in the XX chain, $J=1.0$, we can evaluate the LR velocity as follows:
\begin{equation}
	v_{\mathrm{LR}} \;\approx\; 4 \times 2.71828 \;=\; 10.8731.
\end{equation}
In order to calculate the effective velocity ($v_{\rm eff}$) numerically, we quantify the onset of correlations via the asymmetric QCMI,
\begin{equation}
	I(A;B|C) = \sum_{m} p(m) \,
	I\big( B : C \big)_{\rho_{BC|m}},
\end{equation}
where $A$ is the sender qubit, $B$ is the receiver at distance $d=|A-B|$, $C$ is the set of intermediate sites, $\{p(m)\}$ are probabilities for measurement outcomes on $A$, and $I(B:C)$ is the standard quantum mutual information in the post--measurement state $\rho_{BC|m}$. 
We initialize the chain in the local excitation $\ket{10\cdots0}$, perform a projective measurement on $A$ at each time step, and compute $I(A;B|C;t)$. The \emph{arrival time} $t_\mathrm{arr}(d)$ is defined as the first $t$ where $I(A;B|C;t)$ exceeds a chosen threshold (in Fig.~\ref{fig:effect}, it is $0.03$ bits). The effective velocity is then estimated from the slope of a linear fit,
\begin{equation}
	t_\mathrm{arr}(d) \approx m\, d + b,
\end{equation}
which further yields
\begin{equation}
	\label{eq:slope}
	v_{\mathrm{eff}} \equiv \frac{1}{m}.
\end{equation}
Figure~\ref{fig:effect} depicts the variation of distance $d=\text{dist}(A,B)$ with arrival time $t_\mathrm{arr}(d)$ for $N=8$, $J=1.0$, and $h=0.3$. From the linear fitting, we have obtained the fitting parameter $m \approx 0.397$ and $b \approx -0.504$, which further leads 
\begin{equation}
	v_{\mathrm{eff}} \approx 2.518.
\end{equation}
The calculated values of $v_{\mathrm{LR}}$ and $v_{\mathrm{eff}}$ support the relation: $v_{\rm eff} < v_{\mathrm{LR}}$. Essentially, the LR bound serves as an upper limit. This ensures that no information or operator  can propagate faster than $v_{\mathrm{LR}}$.
We can analytically calculate the group velocity for the XX chain (Eq.~\ref{eq:ham_xx}) and compare it with $v_{\rm eff}$. The group velocity is the physically relevant front speed in free--fermion dynamics and typically controls the observed propagation of many correlation measures. For the XX chain, the Jordan--Wigner transformation~\cite{jordan28} maps the Hamiltonian to free fermions with single--particle dispersion relation:
\begin{equation}
	\varepsilon(k) = 2J\cos k,
\end{equation}
where $k$ is the quasi-momentum. The corresponding group velocity is
\begin{eqnarray}
	v(k) &=& \frac{\partial \varepsilon(k)}{\partial k} \nonumber \\ &=& -2J\sin k.
\end{eqnarray}
Its maximum magnitude occurs at $k = \pi/2$ and equals
\begin{equation}
	v_{0} = 2|J|.
\end{equation}
For $J=1.0$ we therefore have $v_{0} = 2.0$. A discrepancy between $v_0$ and $v_{\mathrm{eff}}$ is visible. Discrete time sampling and small $N$ value, while calculating $v_{\mathrm{eff}}$, are possible sources of this discrepancy. A more robust estimate can be obtained by using finer time resolution, interpolating arrival times, sweeping thresholds, and fitting over larger $N$ with more distances. In the limit of large $N$ and high resolution, $v_{\mathrm{eff}}$ should approach $v_{0}$, in agreement with the LR bound.
\section{Conclusion}
\label{sec:con}
Classical statistics has considerably advanced through the contributions of causal inference. This comprehension enables researchers to derive conclusions about the underlying causal structure exclusively from uncontrolled statistical data, making it a powerful tool applicable across various disciplines of science. Notably, certain paradoxical features of classical correlations have been effectively resolved when analyzed from a causal perspective. This raises an intriguing question regarding the applicability of similar methodologies to quantum correlations. We initiate this exploration by considering the CMI as a causal index that quantum systems may incorporate to scrutinize the causal inference. Consequently, we have developed a rigorous quantum generalization of conditional dependence known as QCMI. After defining QCMI in the framework of quantum interventions, this causal index has been utilized to explore and analyze causal structures in quantum dynamical systems. For example, we have applied it to spin chains, using a projective measurement on one site as the intervention and monitoring its effect on a distant site conditioned on intermediate spins. The results reveal finite-speed propagation of causal influence and coherent oscillations. Subsequently, we have focused on the effective causal propagation velocity, which reflects the speed at which QCMI becomes significant at distant sites.
As a final point, this letter has significantly contributed to this emerging field by proposing a generalized causal index for quantum systems. It discusses the theoretical foundations of this index and emphasizes the applicability of investigating causal measures in quantum many-body systems.

\section*{Acknowledgment}
The author sincerely thanks Dr. M. Palu\v{s} for several fruitful discussions on classical conditional mutual information in the context of causal analysis. The author would also like to extend his sincere appreciation to Dr. N. Ganguly, Dr. S. Das, and Dr. S. Sur for their valuable comments, which have significantly enhanced the quality of this letter.

\appendix

\section{Classical Conditional Mutual Information (CCMI)}
\label{sec:ccmi}
In classical information theory and time series analysis, the CCMI is often used to measure the dependency between two random variables given the knowledge of others. It is defined as follows~\cite{cover06}:
\begin{equation}
	I_c(A:B|C) = H(AC) + H(BC) - H(C) - H(ABC),
\end{equation}
or equivalently:
\begin{equation}
	\label{eq:ccmi}
	I_c(A: B|C) = H(A|C) - H(A|BC),
\end{equation}
where $H(\cdot)$ denotes the Shannon entropy~\cite{cover06,ghosh22,ghosh22_epjp} of the random variables, and the subscript `$c$' implies that the analysis is restricted to the classical domain.
Palu\v{s} et al.~\cite{palus01} reported a specialized form of CCMI, $I(A; B|C)$, which is characterized by its asymmetry concerning variables $A$ and $B$, enabling the measurement of information flow direction from $A$ to $B$ when conditioned on the third variable $C$. Subsequently, a revised version~\cite{palus07} of CCMI, which is an updated and higher-dimensional formulation, has been reported. Let us consider two variables $\{ x_j\}_{j = 1}^{N}$ and $\{ x'_j\}_{j = 1}^{N}$. This latest formulation is frequently utilized in causality studies and is given by: 
\begin{equation}
	\label{eq:ccmi_final}
	I_c(A; B| C) = I_c(x_j ; x'_{j+\tau} \mid \{x'_j, x'_{j-\eta_0}, x'_{j-2\eta_0}\}),
\end{equation}
where $A = x_j$, $B = x'_{j + \tau}$, and $C = \{x'_j, x'_{j-\eta_0}, x'_{j -2\eta_0} \}$. Additionally, $\tau$ (a positive scalar) represents time step; hence,  $x'_{j + \tau}$ is the future state. The other parameter $\eta_0$ is the embedding delay associated with $\{ x'_j\}_{j = 1}^{N}$. In contrast to Eq.~\ref{eq:ccmi}, Eq.~\ref{eq:ccmi_final} demonstrates asymmetry between the variables $A$ and $B$. As a result, we have adopted a different notation, $I_c(A;B|C)$, in Eq.~\ref{eq:ccmi_final} --- rather than using $I_c(A:B|C)$. Intuitively, this particular form of CCMI (Eq.~\ref{eq:ccmi_final}) is asymmetric in $A$ and $B$, and it quantifies the amount of information transferred from the variable $x_j$ to the future value $x'_{j+\tau}$, conditioned on the present and past values of $x'$. Note that Eq.~\ref{eq:ccmi_final} assumes well-defined time-indexed random variables drawn from observable sequences or probability distributions and is typically used in classical settings, such as time-delay embedding or information flow between coupled systems. 
\section{Classical reduction of the QCMI}
\label{sec:reduction}
\ag{When the tripartite quantum state $\rho_{ABC}$ is diagonal in a product basis $\{\ket{abc}\}$, it represents a purely classical joint probability distribution
$p(a,b,c)$:
\begin{equation}
\rho_{ABC} = \sum_{a,b,c} p(a,b,c)\, \ket{abc}\!\bra{abc},
\end{equation}
where $\ket{abc} = \ket{a}_A \otimes \ket{b}_B \otimes \ket{c}_C$. If the local instrument $\mathcal{M}_A$ performs a projective measurement in the same basis,
$\mathcal{M}_A^a(\rho_A) = \ket{a}\!\bra{a}\,\rho_A\,\ket{a}\!\bra{a}$, the measurement merely reads out the classical variable $a$ without introducing any quantum back-action. The post-measurement conditional states are then diagonal,
\begin{equation}
\rho_{BC}^{(a)} = \sum_{b,c} p(b,c|a)\, \ket{bc}\!\bra{bc},
\end{equation}
and their von Neumann entropies reduce to the corresponding Shannon entropies.
Consequently, the asymmetric QCMI
\begin{equation}
I(A;B|C) = \sum_a p(a)\!\left[S(\rho_B^{(a)})+S(\rho_C^{(a)})-S(\rho_{BC}^{(a)})\right]
\end{equation}
reduces exactly to the classical asymmetric CMI computed from $p(a,b,c)$,
\begin{equation}
I(A;B|C) \longrightarrow I_c(x_j; x'_{j+\tau} \mid \{x'_j, x'_{j-\eta_0}, x'_{j-2\eta_0}\}),
\end{equation}
where the variables $A,B,C$ correspond respectively to the source variable $x_j$, the future target $x'_{j+\tau}$, and the conditioning set consisting of the past history $\{x'_j, x'_{j-\eta_0}, x'_{j-2\eta_0}\}$. This limit shows that the QCMI $ I(A;B|C) $ is a natural generalization of the classical asymmetric CMI used to quantify directed information flow and causality in dynamical systems.}
\section{Lieb--Robinson velocity for the XX chain}
\label{sec:lr}
Here, our goal is to find out an explicit (conservative) Lieb--Robinson (LR) velocity~\cite{lieb72} $v_{\mathrm{LR}}$, i.e., a constant $v$ appearing in a bound of the form
\begin{equation}
	\|[A_X(t),B_Y]\| \le C \, \|A_X\|\,\|B_Y\| \, e^{-\mu\big(d(X,Y)-v t\big)},
\end{equation}
for local observables \(A_X,B_Y\) supported on regions \(X,Y\). Now, we shall estimate \(v_{\mathrm{LR}}\) for the short--range XX Hamiltonian (Eq.~\ref{eq:ham_xx}).
Initially, we write the Hamiltonian as a sum of local terms
\begin{equation}
	H=\sum_{Z} h_Z,
\end{equation}
where each two--site interaction across the bond \((i,i+1)\) is
\begin{equation}
	h_{i,i+1} \;=\; \frac{J}{2}\big(\sigma_x^{(i)}\sigma_x^{(i+1)}+\sigma_y^{(i)}\sigma_y^{(i+1)}\big),
\end{equation}
and the single-site field term is \(h_i = h\,\sigma_z^{(i)}\). We need operator-norm bounds for these terms. Using \(\|\sigma_\alpha\|=1\) and submultiplicativity,
\begin{equation}
	\big\|\sigma_x^{(i)}\sigma_x^{(i+1)}\big\| = \|\sigma_x^{(i)}\|\,\|\sigma_x^{(i+1)}\| = 1,
\end{equation}
and likewise for the \(\sigma_y\sigma_y\) product. Hence
\begin{equation}
	\|h_{i,i+1}\| \le \frac{|J|}{2}\big(\,1+1\,\big) = |J|.
\end{equation}
The on-site field satisfies \(\|h_i\| = |h|\,\|\sigma_z\| = |h|\), but single-site terms do not generate propagation between distinct sites and therefore do not increase the effective LR speed beyond the contribution from the two-site couplings.

A common ingredient in LR-style bounds is the quantity
\begin{equation}
	g := \sup_{i}\sum_{Z\ni i} \|h_Z\|,
\end{equation}
the total norm of interactions that include a given site \(i\). For our nearest--neighbor chain each site (in the bulk) participates in two bond terms \(h_{i-1,i}\) and \(h_{i,i+1}\), each of norm \(\le |J|\), plus the on-site field of norm \(|h|\). Thus
\begin{equation}
	g \le 2|J| + |h|.
\end{equation}
Since the field term does not mediate coupling between different sites, it is customary (and slightly tighter) to take the propagation-generating contribution as
\begin{equation}
	g_{\text{prop}} \;\lesssim\; 2|J|.
\end{equation}
There are multiple variants of the LR bound in the literature with different constants. A simple and widely used explicit bound yields an LR velocity proportional to the interaction strength; in one convenient formulation one can take
\begin{equation}
	v_{\mathrm{LR}} \;=\; 2 e \cdot g_{\text{prop}},
\end{equation}
where the factor \(2e\) arises from summing a time series and bounding combinatorial growth by exponentials (this choice is conservative but explicit). Substituting $g_{\text{prop}}\lesssim 2|J|$ gives
\begin{equation}
	\,v_{\mathrm{LR}} \;\lesssim\; 2e\cdot (2|J|) \;=\; 4 e\,|J|\,.
\end{equation}
\bibliographystyle{unsrt}
\bibliography{Ghosh_MS.bib}
\end{document}